\documentclass{article}

\usepackage{amsmath}
\usepackage{amssymb}
\usepackage{cite}

\def\o{\omega}

\def\q{\qquad}
\def\e{\emptyset}

\def\r{\rho}

\begin{document}
\begin{titlepage}
\noindent{\large\textbf{ Multi shocks in reaction-diffusion
models}}

\vskip 2 cm

\begin{center}{Maryam~Arabsalmani{\footnote
{m\_arabsalmani@alzahra.ac.ir}} \& Amir~Aghamohammadi{\footnote
{mohamadi@alzahra.ac.ir}}  } \vskip 5 mm \textit{ Department of
Physics, Alzahra University,
             Tehran 19938-91167, Iran. }

\end{center}

\begin{abstract}
\noindent  It is shown, concerning equivalent classes, that on a
one-dimensional lattice with nearest neighbor interaction, there
are only four independent models possessing double-shocks.
Evolution of the width of the double-shocks in different models is
investigated. Double-shocks may vanish, and the final state is a
state with no shock. There is a model for which at large times the
average width of double-shocks will become smaller. Although there
may exist stationary single-shocks in nearest neighbor reaction
diffusion models, it is seen that in none of these models, there
exist any stationary double-shocks. Models admitting multi-shocks
are classified, and the large time behavior of multi-shock
solutions is also investigated.
\end{abstract}
\begin{center} {\textbf{PACS numbers:}}   02.50.Ga

\end{center}
\end{titlepage}
\section{Introduction}
Recently, shocks in one-dimensional reaction-diffusion models have
absorbed much interest
\cite{PS0,KJS,BS,DLS,PS,J1,PS2,JGM,J2,JM,RS,PRWKGS,MB1,PFF,
RPS,EJS,Bh,AA}. There are some exact results on shocks in
one-dimensional reaction-diffusion models as well as simulations,
numeric results \cite{J1} and also mean field results \cite{PS0}.
Formation of localized shocks in one-dimensional driven diffusive
systems with spacially homogeneous creation and annihilation of
particles has been studied in \cite{PRWKGS}. Recently, in
\cite{KJS},  the families of models with travelling wave solutions
on a finite lattice have been presented. These models are the
Asymmetric Simple Exclusion Process (ASEP), the Branching-
Coalescing Random Walk (BCRW) and the Asymmetric Kawasaki-Glauber
process (AKGP). In all of these cases the time evolution of the
shock measure is equivalent to that of a random walker on a
lattice with $L$ sites with homogeneous hopping rates in the bulk
and special reflection rates at the boundary \cite{KJS}. Shocks
have been studied at both the macroscopic and the microscopic
levels and there are some efforts on addressing the question that
how macroscopic shocks originate from the microscopic dynamics
\cite{PS2}. Hydrodynamic limits are also investigated.

Among the important aspects of reaction-diffusion systems, is the
phase structure of the system. The static phase structure concerns
with the time-independent profiles of the system, while the
dynamical phase structure concerns with the evolution of the
system, specially its relaxation behavior. In
\cite{MA1,AM2,MAM,MA2}, the phase structure of some classes of
single- or multiple-species reaction-diffusion systems have been
investigated. These investigations were based on the one-point
functions of the systems. In a recent article both stationary and
also dynamical single-shocks on a one-dimensional lattice have
been investigated \cite{AA}. It was done for both an infinite
lattice and a finite lattice with boundaries. Static and dynamical
phase transitions of these models have been studied. It was seen
that ASEP has no dynamical phase transition, but both  BCRW and
AKGP have three phases, and the system may show dynamical phase
transitions \cite{AA}.

The question addressed in this article is that, on a
one-dimensional lattice with nearest neighbor interaction, which
models possess double shock and also multi shock solutions. By
double-shock it is meant an uncorrelated state where the
occupation probability has two jumps. All the models have nearest
neighbor interactions and are on a one-dimensional lattice. It is
shown that, concerning equivalent classes, there are only four
independent models possessing double-shocks. For  two models,
double shock disappears and the final state is a linear
combination of Bernoulli measures. There is a model for which At
large times the average width of double shock becomes small.
stationary state is a state which does not evolve. It can be
easily seen that there may exist stationary single shocks in
nearest-neighbor reaction diffusion models (BCRW, and AKGP), in
other words there are single shock states without any evolution.
But in none of these models, there is no stationary double shock.
Combining single shocks one may construct multi shocks. There are
multi shocks of the type $(0,\rho,0,\rho,\cdots)$ and
$(0,1,0,1,\cdots)$. At large times the final state is a linear
combination of single shocks, or a state with no shock.

\section{Notation}
Consider a one-dimensional lattice, each point of which either is
empty or contains one particle. Let the lattice have $L$ sites. An
empty state  is denoted by $|0\rangle $ and an occupied state is
denoted by $|1\rangle $.
\begin{equation}\label{f1}
|0\rangle : = \begin{pmatrix}1\cr 0\end{pmatrix}, \qquad |1\rangle
: = \begin{pmatrix}0\cr 1\end{pmatrix}.
\end{equation}
If the probability that the site $i$ is occupied is  $\rho_i$ then
the state of that  is represented by $\begin{pmatrix}1-\rho_i\cr
\rho_i\end{pmatrix}$.
The state of the system is characterized by
a vector
\begin{equation}\label{f2}
{|\mathbb P\rangle }\in\underbrace{{\mathbb
V}\otimes\cdots\otimes{\mathbb V}}_{L},
\end{equation}
where ${\mathbb V}$ is a $2$-dimensional vector space. All the
elements of the vector ${|\mathbb P\rangle }$ are nonnegative, and
\begin{equation}\label{f3}
{\langle\mathbb S| }{\mathbb P\rangle }=1.
\end{equation}
Here ${\langle\mathbb S| }$ is the tensor-product of $L$ covectors
${\langle s| }$, where ${\langle s| }$ is a covector the
components of which ($s_\alpha$'s) are all equal to one. The
evolution of the state of the system is given by
\begin{equation}\label{f4}
\dot{|\mathbb P\rangle }={\mathcal H}\;{|\mathbb P\rangle },
\end{equation}
where the Hamiltonian ${\mathcal H}$ is stochastic, by which it is
meant that its nondiagonal elements are nonnegative and
\begin{equation}\label{f5}
{\langle\mathbb S| }\; {\mathcal H}=0.
\end{equation}
The interaction is nearest-neighbor, if the Hamiltonian is of the
form
\begin{equation}\label{f6}
{\mathcal H}=\sum_{i=1}^{L-1}H_{i,i+1},
\end{equation}
where
\begin{equation}\label{f7}
H_{i,i+1}:=\underbrace{1\otimes\cdots\otimes 1}_{i-1}\otimes H
\otimes\underbrace{1\otimes\cdots\otimes 1}_{L-1-i}.
\end{equation}
Nondiagonal elements of $H$, shown by $\o_{ij}$, are reaction
rates, hence nonnegative, and  its diagonal elements are
nonpositive. $\o_{ij}$ is the rate for changes of the
configuration of a pair of neighboring sites from the initial
state $j$ to the final state $i$. We take the state $|00\rangle$
as the state $1$, $|01\rangle$ as $2$, $|10\rangle$ as $3$ and
finally $|11\rangle$ as the fourth state. So, e.g. $\o_{23}$ is
the rate for change of configuration $|10\rangle$ to $|01\rangle$,
which is the hoping rate to the right.

 Any configuration of the system may be represented by
the vector ${|E_a\rangle }$. So the system is spanned by $2^L$
vectors, ${|E_a\rangle }$ ($a=1,2,\cdots 2^L$), and any physical
state is a linear combination of these vectors
\begin{equation}\label{f8}
{|\mathbb P\rangle }= \sum_{a=1}^{2^L}{\mathcal P}_a {|E_a\rangle
}, \qquad {\rm where}\qquad \sum_{a=1}^{2^L}{\mathcal P}_a=1.
\end{equation}
${\mathcal P}_a$s are nonnegative real numbers. ${\mathcal P}_a$
is the probability of finding the system in the configuration $a$.

It is said that  the state of the system is a single-shock at the
site $k$ if there is a jump in the density at the site $k$ and the
state of the system is represented by a tensor product of the
states at each site as
\begin{equation}\label{f9}
  |e_k\rangle= u^{\otimes k}\otimes v^{\otimes (L-k)},
\end{equation}
where
\begin{equation}\label{f10}
 u:=\begin{pmatrix}1-\rho_1\cr \rho_1\end{pmatrix} \quad
 v:=\begin{pmatrix}1-\rho_2\cr
 \rho_2\end{pmatrix}.
\end{equation}
It is seen that
\begin{equation}\label{f11}
{\langle\mathbb S| }{e_k\rangle }=1.
\end{equation}
$|e_k\rangle$ represents a state  for which the occupation
probability for the first $k$ sites is $\rho_1$, and  the
occupation probability for the next $L-k$ sites is $\rho_2$. The
set $|e_k\rangle, k=0,1,\cdots L$ is not a complete set, but
linearly independent.

There are three families of stochastic one-dimensional
non-equilibrium lattice models, (ASEP,BCRW,AKGP), for which if the
initial state of these models is a linear superposition of shock
states, at the later times the state of the system ${|\mathbb
P\rangle }$ remains a linear combination of shock states. For
these models
\begin{equation}\label{f12}
    {\mathcal H} |e_k \rangle= d |e_{k-1} \rangle+
    d' |e_{k+1} \rangle-(d+d') |e_{k}\rangle.
\end{equation}
where $d$ and $d'$ are some parameters depending on the reaction
rates in the bulk, and the densities $\rho_1$ and $\rho_2$. So the
span of $|e_k \rangle$'s is an invariant subspace of ${\mathcal
H}$, the Hamiltonian of the above mentioned models. It should be
noted that the number of $|e_k\rangle$'s are $L+1$, and an
arbitrary physical state is not necessarily expressible in terms
of $|e_k \rangle$'s.

Let's assume  that the initial state of the system is a linear
combination of shock states
\begin{equation}\label{f13}
{|\mathbb P\rangle }(0)= \sum_{k=0}^{L}p_k(0) {|e_k\rangle }.
\end{equation}
$p_k$'s, are not necessarily nonnegative, and so any of them  may
be greater than one. For such an initial state, the system remains
in the sub-space spanned by shock measures.
\begin{equation}\label{f14}
{|\mathbb P\rangle }(t)= \sum_{k=0}^{L}p_k(t) {|e_k\rangle }.
\end{equation}
 Using
(\ref{f11}), it is seen that
\begin{equation}\label{f15}
\sum_{k=0}^{L}{p}_k(t)=1.
\end{equation}

 The three models are classified as following \cite{KJS}
\begin{itemize}
    \item 1. {\underline{\rm ASEP}}- The only non-vanishing rates in the bulk
    are the rates of diffusion to the right $\omega_{23}$ and diffusion to the
    left $\omega_{32}$.
    In this case the densities can take any value
    between $0$ and $1$ ($\rho_1,\ \r_2 \ne 0,1$). $d$, and $d'$
    are
\begin{align}\label{f16}
d&=
\frac{\rho_1(1-\rho_1)}{\rho_2-\rho_1}(\omega_{23}-\omega_{32})\nonumber\\
d'&=
\frac{\rho_2(1-\rho_2)}{\rho_2-\rho_1}(\omega_{23}-\omega_{32}).
\end{align}
It should be noted that the densities $\rho_1$, and $\rho_2$ are
also related through
\begin{equation}\label{f17}
 \frac{\rho_2(1-\rho_1)}{\rho_1(1-\rho_2)}=\frac{\omega_{23}}{\omega_{32}}.
\end{equation}
So
\begin{equation}\label{f18}
 d= \frac{\rho_1}{\rho_2}\omega_{23},\qquad d'=
 \frac{\rho_2}{\rho_1}\omega_{32}.
\end{equation}

    \item 2. {\underline{\rm BCRW}}- The non-vanishing rates are coalescence
    ($\omega_{34}$, and $\omega_{24}$), Branching ($\omega_{42}$, and $\omega_{43}$)
    and  diffusion to the left and right ($\omega_{32}$, and $\omega_{23}$).
    The density  $ \rho_1$ can take any value between $0$ and $1$, but $\rho_2$
    should be zero. These parameters are related through
\begin{equation}\label{f19}
\frac{\omega_{23}}{\omega_{43}}=\frac{\omega_{24}+\omega_{34}}{\omega_{42}+\omega_{43}}=\frac{1-\rho_1}{\rho_1}.
\end{equation}

The parameters $d$, and $d'$ are
\begin{align}\label{f20}
d&= (1-\rho_1)\omega_{32}+\rho_1\omega_{34},\cr
d'&=\frac{\omega_{43}}{\rho_1}.
\end{align}
   if  $\omega_{32}=\omega_{34}=\omega_{43}=\omega_{23}=0$, and $\omega_{24}/\omega_{42}=(1-\rho)/\rho$
   then $d=d'=0$, and the model admit stationary single shock.
    \item 3. {\underline {\rm AKGP}}- The non-vanishing rates are
    Death ($\omega_{12}$, and $\omega_{13}$) and Branching to the left and right
    ($\omega_{42}$, and $\omega_{43}$), and also diffusion to the left $\omega_{32}$.
    $\rho_1$ should be equal to one, and $\rho_2$ should be zero.
    The hoping parameters are  $d= \omega_{13}$,
    $d'=\omega_{43}$.

\end{itemize}

\section{Double shocks}
The state of a double shock may be defined through
\begin{equation}\label{d01}
  |e_{m,k}\rangle= u^{\otimes m}\otimes  v^{\otimes k}
  \otimes w^{\otimes
  (L-k-m)},\quad m+k\leq L,
\end{equation}
where
\begin{equation}\label{d02}
 u:=\begin{pmatrix}1-\rho_1\cr \rho_1\end{pmatrix} \quad v:=\begin{pmatrix}1-\rho_2\cr
 \rho_2\end{pmatrix}\quad w:=\begin{pmatrix}1-\rho_3\cr
 \rho_3\end{pmatrix}.
\end{equation}
$|e_{m,k}\rangle$ represents a state for which the occupation
probability for the first $m$ sites is $\rho_1$, the occupation
probability for the next $k$ sites is $\rho_2$, and the occupation
probability for remaining sites is $\rho_3$. We call such state a
double shock, with the first shock at the site $m$, and the other
one at the site $m+k$. $k$ is the width of double-shock, and
$\rho_i\in [0,1]$. To have a double shock $\rho_1$ should be
different from $\rho_2$, and $\rho_2$ also should be different
from $\rho_3$. We search for Hamiltonians for which the span of
$|e_{mk}\rangle$'s is an invariant subspace of ${\mathcal H}$,
\begin{eqnarray}\label{d05}
{\mathcal H} |e_{m,k} \rangle&=& d_1 |e_{m-1,k+1} \rangle+d'_1
|e_{m+1,k-1}\rangle+d_2 |e_{m,k-1} \rangle+
 d'_2 |e_{m,k+1}\rangle\cr &&-(d_1+d'_1+d_2+d'_2) |e_{m,k}\rangle,\qquad k\geq
 2,
\end{eqnarray}
where $d_i$'s ($d'_i$'s) are  parameters depending on the reaction
rates and may be considered as the rates of jump of the shock to
the left (right). ${\mathcal H} |e_{m,1} \rangle$ will be
discussed later.

 As it is seen for single-shocks, one should study
cases with different values of $\r$ separately. One may divide the
region of values for $\r$ to $\r=0$, $0<\r<1$, and $\r=1$. From
now on  the cases $\r=0$, and $\r=1$ will be explicitly stated,
and whenever we write $\rho$, it is meant that $\rho \ne 0,1$. To
have a double-shock there may be different combinations of
densities. There are different models, which may transform to each
other through particle-hole, or right-left interchange. We call
these models equivalent models. As an example the model admitting
the double-shock $(\rho_1,\rho_2,\rho_3)=(0,\rho,1)$ is related to
the model admitting the double-shock $(1,\rho,0)$ through
right-left interchange, and also related to the model admitting
the double shock  $(1,1-\rho,0)$ through particle-hole
interchange. Let's consider the double shock $(0,1,\rho)$. A
necessary condition for the Hamiltonian for which the span of
double-shock measures be an invariant subspace of ${\mathcal H}$,
is that the span of each of single-shock measures $(0,1)$ and
$(1,\rho)$ are separately invariant subspace of ${\mathcal H}$.
The single-shocks $(0,1)$ form an invariant subspace for the
hamiltonian in AKGP. The only interactions which may have nonzero
rates  are
\begin{equation}\label{d06}
    \e A\rightarrow (\e\e, \ AA), \ \
A\e\rightarrow(\e\e,\ AA,\ \e A).
\end{equation}
As far as we consider the single shock $(0,1)$, there is no extra
constraint on the nonzero reaction rates. The single-shocks
$(1,\rho)$ form an invariant subspace for the hamiltonian in BCRW,
with the following interactions
\begin{equation}\label{d07}
    \e \e\rightarrow (\e A, \ A\e), \ \
    \e A\rightarrow(\e\e,\ A\e), \ \
    A \e\rightarrow(\e\e,\ \e A),
\end{equation}
whose reaction rates should satisfy
\begin{equation}\label{d08}
    \frac{\o_{21}+\o_{31}}{\o_{12}+\o_{13}}=\frac{\o_{23}}{\o_{13}}=\frac{\r}{1-\r}.
\end{equation}
The space of parameters of the model, for  a double-shock
$(0,1,\rho)$, is the overlap of the space of parameters of the
AKGP and BCRW. Gathering all these together it is seen that all
the reaction rates should be zero. So there is no reaction
diffusion model with nearest neighbor interaction for which the
double shocks $(0,1,\rho)$ form an invariant subspace.

It can be easily shown that, concerning equivalent classes, there
are only four independent cases.
\begin{itemize}
    \item 1.    $(\rho_1,\rho_2,\rho_3)$,

Among the models possessing shock solution, (and for $\rho_i\ne
0,1$), ASEP is the only model for which double shocks forms an
invariant subspace. The only nonvanishing rates are $\o_{23}$ and
$\o_{32}$, and
    they should satisfy
\begin{equation}\label{09}
\frac{\omega_{23}}{\omega_{32}}=
\frac{\rho_2(1-\rho_1)}{\rho_1(1-\rho_2)}=\frac{\rho_3(1-\rho_2)}{\rho_2(1-\rho_3)}.
\end{equation}
$d_i$'s and $d'_i$'s  are
\begin{equation}\label{100}
    d_1\frac{\rho_2}{\rho_1}=d_2\frac{\rho_3}{\rho_2}=\o_{23}\qquad
    d'_1\frac{\rho_1}{\rho_2}=d'_2\frac{\rho_2}{\rho_3}=\o_{32}
\end{equation}
This model has been studied in \cite{AA}. $\o_{23}$ and $\o_{32}$
are positive nonzero rates. So $d_i$'s and $d'_i$'s are also
nonzero. To have stationary double shock ${\mathcal H} |e_{m,k}
\rangle=0$ leading to  $d_i= d'_i=0$, which is unacceptable. So It
is not possible to have stationary double shock in ASEP.
    \item 2.    $(0,\rho,0)$, (or $(\r, 0, \r)$)

The necessary condition for a model possessing double-shocks
$(0,\rho,0)$ (or $(\r, 0, \r)$) is that this model possesses both
single shocks $(0,\rho)$, and $(\rho,0)$. Nonvanishing rates for
such a model are $\o_{23}, \o_{24}, \o_{32}, \o_{34}, \o_{42}$ and
$\o_{43}$. These rates should satisfy
\begin{equation}\label{11}
\frac{\omega_{24}+\o_{34}}{\omega_{42}+\o_{43}}=\frac{\o_{32}}{\o_{42}}=\frac{\o_{23}}{\o_{43}}=\frac{1-\rho}{\rho}.
\end{equation}
$d_i$'s  and $d'_i$'s are
\begin{eqnarray}\label{12}
d_1=\frac{\o_{42}}{\r},&d'_1= (1-\r)\o_{23}+\r \o_{24}\cr &\cr
d'_2=\frac{\omega_{43}}{\r},&d_2=(1-\r)\o_{32}+\r \o_{34}.
\end{eqnarray}
The Hamiltonian with the above mentioned reaction rates also
possesses  the double-shock $(\r,0,\r)$. The only difference is
that  the rate of jump to the left (and right) of the first
double-shock is the rate of jump to the right (and left) for the
second one. To have stationary double shocks $d_i$'s should be
zero, which needs all the rates to be zero. So, there is no
stationary double shock in this model.

    \item 3.    $(0,1,0)$,

Nonvanishing rates are $\o_{13}, \o_{12}, \o_{42}, \o_{43}$. This
model is  an asymmetric generalization of zero temperature Glauber
model.  $d_i$'s and $d'_i$'s are
\begin{eqnarray}\label{d13}
d_1= \o_{42}& d'_1=\o_{12}\cr d_2=\o_{13},&d'_2=\o_{43}.
\end{eqnarray}
To have stationary double shock $d_i$'s and $d'_i$'s should be
zero, which leads to vanishing all the reaction rates. So, this
model does not have any stationary double shock either.

    \item 4.    $(0,\rho,1)$.

    The  only nonvanishing rate is $\o_{23}$. $d_i$'s and $d'_i$'s   are
\begin{eqnarray}\label{d14}
&d_1=0,&d'_1=(1-\r)\o_{23}\cr \cr  &d_2=\r\ \o_{23},&d'_2=0.
\end{eqnarray}
This model does not have any stationary double shock either.
\end{itemize}

If the initial state is a linear combination of double shocks,
then
\begin{equation}\label{d9}
{|\mathbb P\rangle}(t)=
\sum_{m=-\infty}^{\infty}\sum_{k=1}^{\infty}p_{mk}(t)
{|e_{mk}\rangle },
\end{equation}
where $p_{m,k}$ is the contribution of  the double-shock, $(mk)$
in the state of the system.  Using (\ref{f4}), (\ref{d9}), and
also the linear independency of $|e_{m,k}\rangle$'s, one can
obtain the evolution equation for $p_{mk}$'s. It is  a difficult
task to solve difference equations of this type. $p_{m,k}$ has two
indices, $m$ representing the position of the first shock, and $k$
the width of the double shock. We may forget about the position of
the first shock and ask only about the width of double shock. Then
one may encounter with a difference equation which can be solved
more easier.

Let's consider the general case where the span of
$|e_{mk}\rangle$'s is an invariant subspace of ${\mathcal H}$
\begin{equation}\label{d15-1}
    {\mathcal H} |e_{m,k} \rangle=\sum_{m',k'}{\mathcal H}_{mk}^{m'k'}
    |e_{m',k'}\rangle.
\end{equation}
If the Hamiltonian has the property that $\sum_{m'}{\mathcal
H}_{mk}^{m'k'}$ is independent of $m$, then one may define a new
Hamiltonian $\widetilde{\mathcal H}$ through
\begin{equation}\label{d15-2}
\widetilde{\mathcal H}_{k}^{k'}:=\sum_{m'}{\mathcal H}_{mk}^{m'k'}
\end{equation}
It can be easily shown that $\widetilde{\mathcal H}$ is
stochastic, it is meant that
\begin{eqnarray}\label{d15-3}
&\widetilde{\mathcal H}_{k}^{k'}>0,&\qquad {\rm for}\quad k'\ne k
\cr &&\cr &\sum_{k'}\widetilde{\mathcal H}_{k}^{k'}=0.&
\end{eqnarray}
Then one may forget about $m$, position of the first shock, and
only ask about the contribution of double shocks with the width
$k$. It is obvious that some part of information about the
position  of the first shock will be lost. Now one may define
$|f_{k}\rangle $ as the state of a double shock with the width
$k$. Identifying  all $|e_{m,k}\rangle$ with the same $m$ to each
other in the state (\ref{d9}), one may define another state
$\widetilde{|\mathbb P\rangle}$ where the information of the
position of the first shock has being lost
\begin{equation}\label{d15-4}
\widetilde{|\mathbb P\rangle}(t)= \sum_{k=1}^{\infty}q_{ k}(t)
|f_{k}\rangle.
\end{equation}
Here $q_k(t)$ is defined through
\begin{equation}\label{d11}
    q_k:=\sum_{m=-\infty}^{\infty}p_{mk}.
\end{equation}
and it  is the contribution of all double shocks with the width
$k$.  Then instead of (\ref{d05}), one may obtain
\begin{eqnarray}\label{d16}
 \widetilde{\mathcal H} |f_{k} \rangle&=& D |f_{k+1} \rangle+D' |f_{k-1}\rangle
 -(D+D') |f_{k}\rangle,\qquad k\geq 2.
\end{eqnarray}
where
\begin{equation}\label{d13-1}
D:= d_1+d'_2,\qquad D':= d'_1+d_2.
\end{equation}

\subsection{ Double-shocks $(0, \rho,0)$ and $(0,1,0)$ on a periodic lattice}
Let's consider a lattice with $L$ sites and with periodic boundary
conditions. Then, the only double shocks which could exist, are
$(0,1,0)$ and $(0,\r,0)$. Let's sum up the contributions of all
double shocks with the same width.  The position of double shocks
will be again washed out. Then one should work with $|f_k\rangle$,
which stands for the state of a double-shock with the width $k$.
$|f_0\rangle$, and $|f_L\rangle$ are Bernoulli measures
corresponding to an empty lattice and a full lattice,
respectively. It can be easily shown that

\begin{eqnarray}\label{d16-2}
&&\widetilde{\mathcal H}|f_0\rangle=0,\cr &&\widetilde{\mathcal
H}|f_k\rangle=D|f_{k+1}\rangle+D'|f_{k-1}\rangle-(D+D')|f_k\rangle\q
k\neq0,L,\cr &&\widetilde{\mathcal H}|f_L\rangle=0,
\end{eqnarray}
Here $D$ stands for the rate of increasing the width of
double-shock, and $D'$ stands for the rate of decreasing its
width. One can map this model to a model with one particle on
lattice with boundaries at $k=0$, and $k=L$. This particle hops to
the right and left with the rates $D$ and $D'$, and there are
traps at the boundaries. The system has only two stationary state,
$|f_0\rangle$, and $|f_L\rangle$, means that at large times there
is no shock, and the final state is a linear combinations of the
Bernoulli measures.
\begin{equation}\label{d18}
 \widetilde{|\mathbb P\rangle }= q_0|f_0\rangle +q_L|f_L\rangle.
\end{equation}
If the initial state is a linear combination of $|f_k\rangle$'s
then
\begin{equation}\label{d19}
{\widetilde{|\mathbb P\rangle }}(t)=
\sum_{k=0}^{L}q_k(t)|f_k\rangle.
\end{equation}
Using (\ref{d13-1}), one arrives at
\begin{eqnarray}\label{d20}
&&\dot q_0=D'q_1,\cr &&\dot{q_1}=D'q_2-(D+D')q_1,\cr &&\dot
q_k=D'q_{k+1}+Dq_{k-1}-(D+D')q_k\q k\neq0,1,L-1,L,\cr &&\dot
q_{L-1}=Dq_{L-2}-(D+D')q_{L-1},\cr &&\dot{q_L}=Dq_{L-1}.
\end{eqnarray}
$q_k(t)$'s in the bulk, ($k\ne 0,L$), can be obtained. They are
\begin{eqnarray}\label{d21}
q_k(t)&=&\frac{2}{L}(\frac{D}{D'})^{k/2}e^{-(D+D')t}\sum_{s=1}^{L-1}
\sum_{m=1}^{L-1}q_m(0)(\frac{D}{D'})^{m/2}\cr && \sin{(\frac{s\pi
m}{L})}\sin{(\frac{s\pi k}{L})}e^{2t\sqrt{DD'}\cos{(s\pi/L)}}.
\end{eqnarray}
One may integrate $q_1(t)$, and $q_{L-1}(t)$ to obtain $q_0(t)$,
and $q_L(t)$, which are the only terms surviving at large times.
There is also another way to obtain the $q_0$, and $q_L$ at
infinitely large times. In fact, there are two constants of motion
${\cal I}_1$ and ${\cal I}_2$. ${\cal I}_1$ is related to the
conservation of probability
\begin{equation}\label{d22}
{\langle\mathbb S| }{\widetilde{\mathbb P\rangle }}=1\quad
\Rightarrow \quad {\cal I}_1:=\sum_{k=0}^Lq_k(t)=1,
\end{equation}
and
\begin{equation}\label{d23}
   {\cal I}_2:=\sum_{k=0}^Lq_k(t)(\frac{D'}{D})^k.
\end{equation}
It should be noted that the system has  two stationary states, so
there are two right eigenvectors corresponding to zero eigenvalue
for the Hamiltonian ${\mathcal H}$. Therefore there are also two
left eigenvectors corresponding to zero eigenvalue for ${\mathcal
H}$. These are

\begin{equation}\label{d25}
\langle\mathbb S| =\begin{pmatrix}1& 1& 1&\cdots &1\end{pmatrix}.
\end{equation}
\begin{equation}\label{d26}
\langle\mathbb S'|
=\begin{pmatrix}1&\frac{D'}{D}&(\frac{D'}{D})^2&(\frac{D'}{D})^3&\cdots&(\frac{D'}{D})^L\end{pmatrix}.
\end{equation}
The second constant of motion can be obtained using
${\langle\mathbb S'| }{\widetilde{\mathbb P\rangle }}(t)$. As long
as $D\ne D'$, the constants of motion ${\cal I}_1$ and ${\cal
I}_2$ are two independent quantities. For $D=D'$, ${\cal I}_1$ and
${\cal I}_2$ are the same. But as the stationary state has
two-fold degeneracy, there should exist another constant of
motion. The second independent constant of motion is ${\cal
I'}_2:=\sum_{k=0}^Lk\ q_k(t)= \langle k\rangle$. So, for $D=D'$,
the average width of the shock, $\langle k\rangle$, is a constant
of motion. One should expect this, because $D$ and $D'$ are the
rates for increasing the width of the double shock and decreasing
it respectively.

For the double-shock $(0,\r,0)$, $D'/D=1-\r<1$. So, the constant
of motions are ${\cal I}_1$ and ${\cal I}_2$.  ${\cal I}_1$  is
the summation of probabilities for finding a double shock with any
width, so it should be equal to one. The second constant of motion
also has a physical interpretation. The rate for changing any
configuration of a pair of neighboring sites to the state
$|\e\e\rangle$ is zero. So the probability for finding a
completely empty lattice does not change with time. ${\cal I}_2$
is exactly the probability of finding an empty lattice in the
initial state
\begin{equation}\label{d28}
{\cal I}_2=\sum_{k=0}^Lq_k(t)(1-\r)^k=\sum_{k=0}^Lq_k(0)(1-\r)^k.
\end{equation}
Using constant of motions, for $D\ne D'$,  at infinitely large
times, we have
\begin{eqnarray}\label{d24}
&&q_0+q_L=1,\cr\cr
&&q_0+(\frac{D'}{D})^Lq_L=\sum_{k=0}^Lq_k(0)(\frac{D'}{D})^k.
\end{eqnarray}
Solving these equations one obtains
\begin{eqnarray}\label{d24-1}
&&q_0(\infty)=\Big[\sum_{k=0}^Lq_k(0)
(\frac{D'}{D})^k-(\frac{D'}{D})^L\Big]\Big/
\Big[1-(\frac{D'}{D})^L\Big],\cr\cr
&&q_L(\infty)=\Big[1-\sum_{k=0}^Lq_k(0)
(\frac{D'}{D})^k\Big]\Big/\Big[1-(\frac{D'}{D})^L\Big].
\end{eqnarray}
As it is seen the contribution of $|f_0\rangle$ and $|f_L\rangle$
in the final state depends on both reaction rates and initial
conditions.

The Hamiltonian for the model possessing the double-shock $(010)$,
with $D=D'$ is the Hamiltonian for zero temperature Glauber model.
This model have been studied in \cite{AM0,MA1,RAM}. The average
density at each site $\langle n_i\rangle(t) $ at the time $t$ and
also all the correlation functions at large times for an infinite
lattice have been calculated in \cite{AM0}. Static and dynamical
phase transitions of this model have been   also studied in
\cite{MA1}. Here, $D$ is not necessarily equal to $D'$. For
$D'>D$, and large $L$, one arrives at
\begin{eqnarray}\label{d24-2}
&&q_0(\infty)=1-\sum_{k=0}^Lq_k(0)(\frac{D'}{D})^{k-L},\cr\cr
&&q_L(\infty)=1-q_0(\infty),
\end{eqnarray}
If initially only double shocks with finite widths have
contributions, in the thermodynamic limit $(L\to \infty )$ the
system will finally fall in the state $f_0$, but if $D'<D$ it can
be seen that both stationary states have contributions in the
final state.

For the case $D=D'$, one obtains
\begin{eqnarray}\label{d27}
&&q_0(\infty)=1- \frac{1}{L}\sum_{k=0}^Lk\ q_k(0)=1-
\frac{1}{L}\langle k\rangle,\cr\cr &&q_L(\infty)=
\frac{1}{L}\sum_{k=0}^Lk\ q_k(0)= \frac{1}{L}\langle k\rangle,
\end{eqnarray}
which means that at large times the system is fully occupied or
empty. The probability of finding a fully occupied lattice at
large times is equal to the initial average width of the
double-shock divided by the size of the lattice.

\subsection{ Double-shock $(0,\r,1)$ }
Let's consider the double-shock $(0,\r,1)$ on an infinite
lattice. The only non-vanishing rate is $\o_{23}$, which can be
set equal to $1$, by a suitable redefinition of time. Direct
calculation gives

\begin{eqnarray}\label{d28-1}
&&{\mathcal H}|e_{m,1}\rangle =0,\cr &&{\mathcal
H}|e_{m,k}\rangle=(1-\r)|e_{m+1,k-1}\rangle+\r\ |e_{m,k-1}\rangle
-|e_{m,k}\rangle,\q k\neq1.
\end{eqnarray}
It is seen that there is no probability for width increase. If
there is initially a shock $|e_{mk}\rangle$, at later times its
width becomes smaller, and at large times there are only double
shocks with the width 1. Starting with a linear combination of the
shocks, the evolution equation for $p_{mk}$'s, can be obtained to
be
\begin{eqnarray}\label{d30}
&&\dot{p}_{m,1}=(1-\r)p_{m-1,2}+\r p_{m,2},\cr
&&\dot{p}_{m,k}=(1-\r)p_{m-1,k+1}+\r p_{m,k+1}-p_{m,k}\q k\neq1.
\end{eqnarray}
Defining $q_k$, through (\ref{d11}), one arrives at
\begin{eqnarray}\label{d31}
&&\dot{q}_1=q_2,\cr &&\dot{q}_k= q_{k+1}-q_k \q k\neq 1.
\end{eqnarray}

If initially the state of the system is a double shock, e.g. $|
e_{MK}\rangle$. Then, it is obvious that at later times there are
only double shocks with the position of the first shock  in the
range $M\leq m\leq M+K-1$, and with the width $1\leq k\leq M+K-m$.
Let's assume the initial state is
\begin{equation}\label{d32}
{\widetilde{|\mathbb P\rangle}
}=\sum_{k=0}^{L}q_{k}(0)|f_{k}\rangle,
\end{equation}
where $|f_k\rangle$ is the state of double shocks with the width
$k$. Then
\begin{eqnarray}\label{d33}
&&\dot{q}_0=0,\cr &&\dot{q}_1=q_2,\cr &&\dot{q}_k= q_{k+1}-q_k, \q
2\leq k\leq L-1, \cr &&\dot{q}_L=-q_L,\cr &&\dot{q}_k= 0, \q\q\q
L+1\leq k.
\end{eqnarray}
The above equations show that at large times there are only
contributions of the double shocks with the width 1. These set of
equation can be solved  leading to
\begin{equation}\label{d34}
    q_k(t)= \sum_{n=0}^{L-k}q_{k+n}(0)\frac{t^n}{n!}\ {\rm e}^{-t}\q
    2\leq k \leq L.
\end{equation}
This together with $\dot q_1=q_2$ can be used to obtain $q_1(t)$.
\begin{align}\label{d35}
    q_1(t)=&\ q_1(0)+ \sum_{n=0}^{L-2}q_{n+2}(0)\int_0^t\frac{t'^n}{n!}\ {\rm
    e}^{-t'}\nonumber\\
          =&\sum_{n=1}^Lq_n(0)- \sum_{n=0}^{L-2}\sum_{m=0}^n q_{n+2}(0)\frac{t^m}{m!}\ {\rm
    e}^{-t}.
\end{align}
As it is expected at large times all the double shocks changes to
the double-shock with the width 1.

Let's study the distribution of these double shocks at large
times. Using (\ref{d28-1}), and defining $A_{m,k}:=\exp(t{\mathcal
H})|e_{mk}\rangle$, it is seen that

\begin{align}\label{d36}
    \frac{\partial A_{m,k}}{\partial t}+
    A_{m,k}=&(1-\r)A_{m+1,k-1}+\r A_{m,k-1}\q k\ne 1,\nonumber \\
    A_{m,1}=& |e_{k1}\rangle.
\end{align}
At large times this equation recasts to
\begin{equation}\label{d37}
    A_{m,k}(\infty)=(1-\r)A_{m+1,k-1}(\infty)+\r A_{m,k-1}(\infty)\q k\ne 1,
\end{equation}
whose solution is obtained to be
\begin{equation}\label{d38}
A_{m,k}(\infty)=\lim_{t\to \infty}\left({\rm e}^{t{\mathcal
H}}|e_{mk}\rangle\right) =\sum_{j=0}^{k-1}
\begin{pmatrix}k-1\cr j\end{pmatrix}
(1-\r)^j\r^{k-1-j}\ |e_{m+j,1}\rangle.
\end{equation}
So, at large times the state of the system is a linear combination
of double shocks with the width 1. The distribution of the
position of these double shocks is a binomial distribution. Let's
consider the initial state to be a double shock with the width
$k$, $|e_{0,k}\rangle$, then the average position of the first
shock at large times is
\begin{equation}\label{d39}
    \langle j\rangle= (k-1)(1-\r),
\end{equation}
and the width of the binomial distribution is
$\sqrt{\r(1-\r)(k-1)}$.

\section{Multi shocks}
Combining single shocks one may construct multi shocks. The only
models with multi shocks are as following
\begin{itemize}
    \item 1. $(\rho_1,\rho_2,\rho_3,\cdots)$

     The span of multi shocks $(\rho_1,\rho_2,\rho_3,\cdots)$
     is an invariant subspace of Hamiltonian of ASEP
     provided the densities satisfy
     \begin{equation}\label{ms1}
           \frac{\rho_{i+1}(1-\rho_i)}{\rho_i(1-\rho_{i+1})}=\frac{\omega_{23}}{\omega_{32}}.
     \end{equation}
     It can be easily seen that the
     rate of hoping of the $i$th shock to the left, $d_i$, and also the
     rate of hoping of the $i$th shock to the right, $d'_i$, is
     given by
     \begin{align}\label{ms2}
d_i=&\o_{23}\frac{\rho_i}{\rho_{i+1}}\nonumber\\
d'_i=&\o_{32}\frac{\rho_{i+1}}{\rho_i}.
\end{align}
This result are   obtained in \cite{BS}.
    \item 2.$(0,\rho,0,\rho,\cdots)$ and $(0,1,0,1,\cdots)$

The model admitting multi shock of the type
$(0,\rho,0,\rho,\cdots)$ are that admitting double shocks
$(0,\rho,0)$, or $(\rho,0,\rho)$. The model Possessing multi
shocks of the type $(0,1,0,1,\cdots)$ is an asymmetric
generalization of the zero temperature Glauber model. In both of
these multi shocks there are edges at shock points.   The edges
are destroyed two by two. Let's consider a multi shock of order
$N$ with the first shock at the site $m$. It is seen that the
action of Hamiltonian on such state is
\begin{eqnarray}\label{ms4}
{\mathcal H} |e_{m,k_1,\cdots, k_{N-1}} \rangle&=& d_1
|e_{m-1,k_1+1,\cdots, k_{N-1}}\rangle +d'_1 |e_{m+1,k_1-1,\cdots,
k_{N-1}}\rangle\cr &&\cr &&+d_2 |e_{m,k_1-1,\cdots, k_{N-1}}
\rangle+
 d'_2 |e_{m,k_1+1,\cdots, k_{N-1}}\rangle+\cdots \cr &&\cr &&+d_2
|e_{m,k_1,\cdots, k_{N-1}-1}\rangle +d'_2 |e_{m,k_1,\cdots,
k_{N-1}+1}\rangle.
\end{eqnarray}
If any of the $k_i$s in the left hand side is equal to one, at the
right hand side there will be a multi shock of order $N-2$. So
there is  a finite probability that the system falls in a state
with lower shocks, and there is no probability for increasing the
number of shocks. In fact if there exists a state for which  any
state can transform directly or even indirectly to it, and that
state has no evolution, then that state is the final stationary
state.  Let's consider a periodic lattice. The number of shocks,
$N$, should be even.  So at large times the state of system is  a
state with no shock. For the models on an infinite lattice number
of shocks,$N$,  may be  even or odd. Then  For odd $N$  the final
state is a linear combination of single shocks.

\end{itemize}

\section{Summary}
There are three types of models with travelling wave solutions on
a one-dimensional lattice. These are classified in \cite{KJS}. It
is seen that there are four type of models admitting double
shocks. Double shocks and the models admitting these double shocks
are as following
\begin{itemize}
\item $(\rho_1,\rho_2,\rho_3)$.
Nonvanishing rates are $\omega_{23}$, and $\omega_{32}$.
\item  $(0,\rho,0)$, (and also $(\rho,0,\rho)$). Nonvanishing rates are
$\o_{23}, \o_{24}, \o_{32}, \o_{34}, \o_{42}$ and $\o_{43}$.
\item $(0,1,0)$. Nonvanishing rates are $\o_{13}, \o_{12}, \o_{42}, \o_{43}$.
\item $(0,\rho,1)$. The only nonvanishing rate is is $\o_{23}$.
\end{itemize}

There are three type of models admitting multi shocks. The multi
shocks are of the type
\begin{itemize}
\item $(\rho_1,\rho_2,\rho_3,\cdots)$. Nonvanishing rates are $\omega_{23}$, and $\omega_{32}$.
\item $(0,\rho,0,\rho,\cdots)$. Nonvanishing rates are $\o_{23}, \o_{24}, \o_{32}, \o_{34}, \o_{42}$ and $\o_{43}$.
\item  $(0,1,0,1,\cdots)$. Nonvanishing rates are $\o_{13}, \o_{12}, \o_{42}, \o_{43}$.
\end{itemize}

{\bf Acknowledgment} This work was partially supported by the
research council of the Alzahra University. We would like to thank
M. Khorrami for useful discussions.

\end{document}